\documentclass[letterpaper,twocolumn,10pt]{article}
\usepackage{wrapfig}

\usepackage{algorithm}
\usepackage{algorithmic}
\usepackage{amssymb}
\usepackage{xspace}
\usepackage{comment}
\usepackage{cite}
\usepackage{subfigure}

\newcommand{\algorithmicinput}{\textbf{Input:}}
\newcommand{\INPUT}{\item[\algorithmicinput]}

\usepackage{graphicx}
\usepackage{color}

\newcommand{\deps}{$\ \textsf{dep} \ $}

\newcommand{\name}{ACCF\xspace}

\usepackage{todonotes}
\usepackage{url}

\usepackage{usenix,epsfig,endnotes}
\begin{document}

\date{}

\title{\Large \bf Toward Adaptive Causal Consistency for Replicated Data Stores}

\author{
{\rm Mohammad \ Roohitavaf}\\
Michigan State University
\and
{\rm Sandeep \ Kulkarni}\\
Michigan State University
} 

\maketitle
\begin{abstract}

Causal consistency for key-value stores has two main requirements (1) do not make a version visible if some of its dependencies are invisible as it may violate causal consistency in the future and (2) make a version visible as soon as possible so that clients have the most recent information (to the extent feasible). 
These two requirements conflict with each other. 
Existing key-value stores that provide causal consistency (or detection of causal violation) utilize a static approach in the trade-off between these requirements. Depending upon the choice, it assists some applications and penalizes some applications. 
We propose an alternative where the system provides a set of tracking groups and checking groups. This allows the application to choose the settings that are most suitable for that application. Furthermore, these groups can be dynamically changed based on application requirements.

\end{abstract}
\newcommand{\br}[1]{\ensuremath{\langle #1 \rangle}\xspace}
\section{Introduction}
\label{sec:intro}
\begin{figure}[htp]
  \centering
  \subfigure[$2\times2$ \label{fig:full}]
  {\includegraphics[scale=0.35]{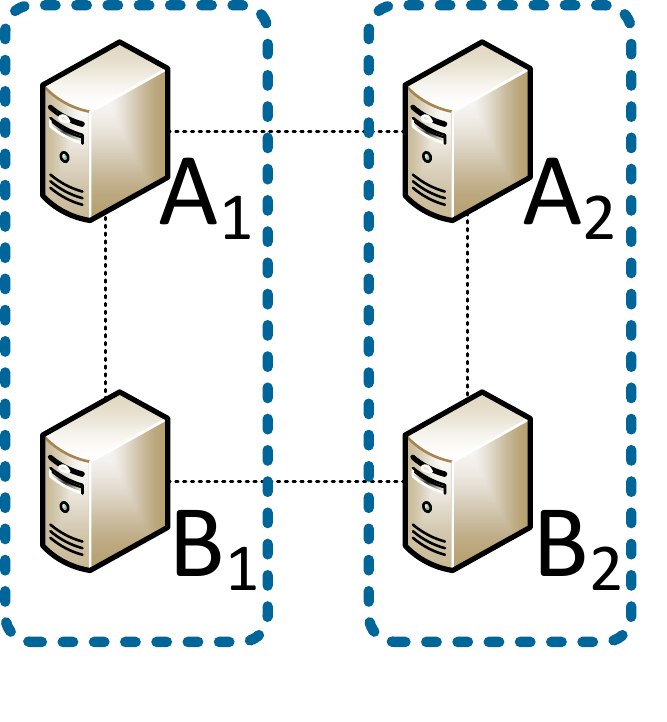}} \quad
  \subfigure[$4\times1$\label{fig:partial}]
  {\includegraphics[scale=0.35]{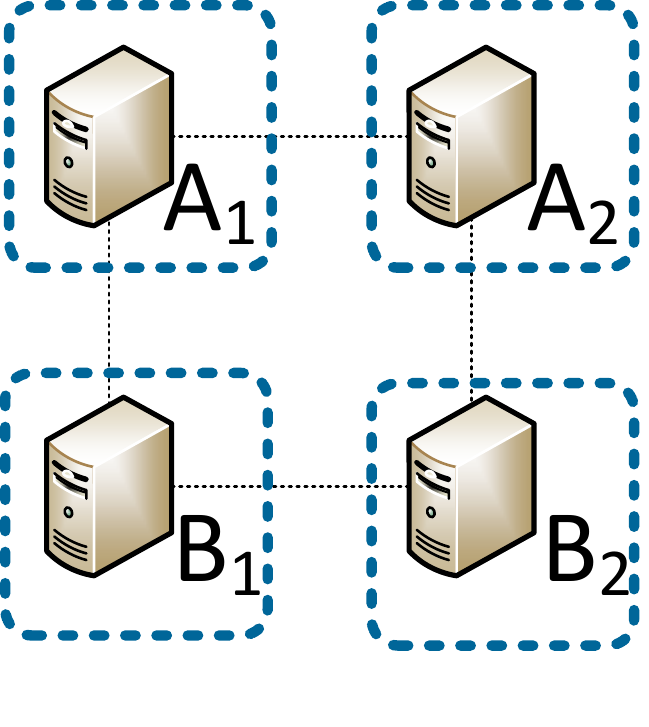}}
 \caption{Two ways to organize replicas}
 \label{fig:introFig}
\end{figure}

\vspace*{-2mm}
Causal consistency for distributed key-value stores has received much attention from academia in recent years. 
Existing protocols utilize a static approach in the trade-off between different conflicting requirements (e.g. consistency, visibility, and throughput).  They also treat all clients the same, and assume that their usage patterns are always unchanged. For example, they assume clients only access their local data center, and any client may access any part of the data. However, different applications may have different usage patterns.
To illustrate, consider a simple system that consists of two partitions $A$ and $B$ with geographically distributed copies $A_1$, $A_2$, $B_1$ and $B_2$. Suppose, we are using a causal consistency protocol like \cite{cops, eiger, orbe, gentleRain, causalSpartan} that does not make a version visible, unless it made sure all partitions inside a replica are updated enough. 
We consider two possible ways to organize the replicas: (1) two full replicas each with two partitions, referred to as $2\times2$ (2) or four partial replicas each with one partition referred to as $4\times1$. These two organizations are shown in Figure \ref{fig:introFig}. Now, consider two applications. The first application, $App_1$ consists of two clients $C1$ and $C2$ that access $A_1$ and $A_2$ respectively for a collaborative work. In $App_1$, each client updates the data after it reads the new version written by another client. 
Since each client waits for the other client's update, any increase in update visibility will reduce the throughput of $App_1$. In the scenario in Figure \ref{fig:full}, since $A_1$ and $B_1$ are considered in the same replica, $A_1$ does not make versions visible, unless it made sure $B_1$ is updated enough. Thus, if the communication between $A_1$ and $B_1$ is slow, it takes more time for $A_1$ to make a version visible. Since the data on $B_1$ and $B_2$ is irrelevant for $App_1$, this delay by $A_1$ is unnecessary which leads to increased visibility latency which, in turn, leads to a reduced throughput of $App_1$. Furthermore, if there were a large number of such partitions, this delay would be even more pronounced. 
%
By contrast, there is no such penalty in scenario in Figure \ref{fig:partial}, as in Figure \ref{fig:partial}, partitions $A_1$ and $B_1$ are considered in different replicas. Thus, they do not check each other. 

On the other hand, consider $App_2$ that consists of one client, say $C3$, and it accesses data from $A_1$ and $B_1$. In scenario in Figure \ref{fig:full}, $C_3$ is guaranteed to always read the consistent data. However, in scenario in Figure \ref{fig:partial}, since $A_1$ and $B_1$ do not check the freshness of each other, $C_3$ may suffer from finding inconsistent versions (or delays or repeated requests to find a consistent version) while accessing $A_1$ and $B_1$. 

From the above discussion, it follows that no matter how we configure the given key-value store, a system with a static configuration that treats all clients the same will penalize some clients. 
%
%
%
%
The goal of this work is to develop a broad framework that instead of relying on a fixed set of assumptions, allows the system to be dynamically reconfigured after learning the actual client activities and requirements.

In Section \ref{sec:approach}, we propose an approach that lets us effectively trade off between different objectives and serve different groups of clients differently. Next, in Section \ref{sec:protocol}, we provide a framework that uses our proposed approach. In Section \ref{sec:discussion}, we discuss ideas for creating adaptive causal systems based on our protocol. Finally, in Section \ref{sec:con} we conclude the paper.

\section{Adaptive Causal Consistency}
\label{sec:approach}
The broad approach for providing causal consistency is to track the causal dependencies of a version, and check them before making the version visible in another replica.
%
%
%
%
Tracking and checking are usually done using timestamping versions as follows: 

\begin{itemize}
\item \textit{Dependency Tracking}: Upon creating a new version for a key, we assign a timestamp to the version that \textit{somehow} captures causal dependencies of the version. 

\item \textit{Dependency Checking}: 
Upon receiving a version, the receiving replica does not make the version visible to the clients until it makes sure that all of the dependencies of the version are also visible to the clients. 
\end{itemize}

The goal of timestamping is to provide a way to capture causal relation between two versions. To satisfy $v_1 \deps v_2 \Leftrightarrow v_1.t > v_2.t$ (where $v_1 \deps v_2$ means the event of writing $v_2$ has happened-before \cite{lamport} the event of writing $v_1$, and $v.t$ is the timestamps assigned to $v$), we need timestamps of size $O(N)$ \cite{bost} where $N$ is number of nodes that clients can write on.
To solve the issue of large timestamps, causal consistency protocols consider servers in groups and track causality with vectors that have one entry per group. We refer to such groups as \textit{tracking groups}. Tracking dependencies in groups, provides timestamps that satisfies a weaker condition  $v_1 \deps v_2 \Rightarrow v_1.t > v_2.t$. This condition lets us guarantee causal visibility of the versions. However, since it does not provide accurate causality information, we may need to unnecessarily delay the visibility of a version by waiting for versions that are not its real dependencies. Thus, by grouping servers in tracking group, we trade off the visibility of versions for a lower metadata size. 

We face a similar trade-off in the dependency checking. Dependency checking determines how conservative we are in making versions visible to the clients. 
Since checking the whole system is expensive, causal consistency protocols consider systems in groups, and each server only checks servers in its own group. We refer to such groups as \textit{checking groups}. Most of current protocols \cite{cops,eiger, orbe,gentleRain,gentleRain+, causalSpartan} group servers by their replicas. Thus, a server only checks the dependencies inside the replica that it belongs to.
Table \ref{tab:protocols} shows tracking and checking groups for some of the recent causal systems.

\begin{table}[]
\centering
\caption{Tracking and Checking in Some of Causal Systems}
\label{tab:protocols}
{\footnotesize
\begin{tabular}{lllll}
\textbf{Protocol} & \textbf{Tracking} & \textbf{Checking} \\ 
COPS \cite{cops}             & Per key        & Per Replica       \\ 
Eiger \cite{eiger}             & Per key        & Per Replica     \\ 
Orbe \cite{orbe}             & Per server        & Per Replica      \\ 
GentleRain   \cite{gentleRain}     & Per system    & Per Replica     \\ 
Occult    \cite{occult}        & Per Master Server & No checking    \\ 
Okapi    \cite{okapi}        & Per Replica  & Per system     \\ 
CausalSpartan \cite{causalSpartan}    & Per Replica       & Per Replica     \\ 

\end{tabular}
}
\end{table}



When we are designing a causally consistent key-value store, two natural questions arise based on the trade-offs explained above: 1) how much tracking accuracy is enough for a system? 
2) how much should we be conservative in making versions visible? 
We believe the answer to these questions depends on the factors that should be learned at the run-time. A practical distributed data store performs in a constantly changing environment; the usage pattern of clients can change due to many reasons including time of the day in different time zones or changes in load balancing policies; data distribution can change, because we may need to add or remove some replicas; components may fail or slow down, and so on. These changes can easily invalidate assumptions made by existing causal consistency protocols such as \cite{cops,eiger, orbe,gentleRain,causalSpartan,occult} which leads to their reduced performance in practical settings \cite{facebook}. To solve this issue, we believe that a key-value store must monitor the factors  mentioned above and \textit{dynamically} trade off between different conflicting objectives. We believe dynamically changing tracking and checking grouping based on what we learn from the system is an effective approach to perform such dynamic trade-offs. Using a flexible tracking and checking grouping we are also able to treat different applications in different ways.

To use the above approach, however, we need a protocol that can be easily configured for different groupings. As shown in Table \ref{tab:protocols}, existing protocols assume fixed groupings that cannot be changed. To solve this issue, in the next section, we provide a protocol that can be configured to use any desired grouping. This flexible algorithm provides a basis for creating adaptive causal systems. This algorithm also lets us treat clients in different ways, and unlike most of the existing protocols that require a certain data distribution schema, our algorithm allows us to replicate and partition our data any way we like including creating partial replicas. 
\section{Adaptive Causal Consistency Framework} 
\label{sec:protocol}

In this section, we provide Adaptive Causal Consistency Framework (ACCF) which is a configurable framework that lets us deal with trade-offs explained in Section \ref{sec:approach}. Specifically, as the input, \name receives 1) function $T$ that assigns each server to exactly one tracking group, and 2) function $C$ that assigns each server to a \textit{non-empty set} of checking groups.


\subsection{Client-side}

Algorithm \ref{alg:client} shows the client-side of the \name. A client $c$ maintains a set of pairs of tracking group ids and timestamps called dependency set, denoted as $DS_c$. For each tracking group $i$, there is at most one entry $\langle i, h \rangle$ in $DS_c$ where $h$ specifies the maximum timestamp of versions read by client $c$ originally written in servers of tracking group $i$. 
Each data object has a key and a version chain containing different versions for the object. Each version is a tuple $\langle v, ds \rangle$, where $v$ is the value of the version, and $ds$ is a list that has at most one entry per tracking group that capture dependency of the version on writes on different tracking groups.

\begin{algorithm}[h]
{
\caption{Client operations at client $c$}
\label{alg:client}
\begin{algorithmic} [1]
{\footnotesize
\INPUT Load balancer $L$
\STATE \textbf{GET} (key $k$, checking group id $cg$) 
\STATE \hspace{3mm} $i = L(k)$ 
\STATE \hspace{3mm}  send $\langle \textsc{GetReq} \ k, cg, DS_c\rangle$  to server $i$
\STATE \hspace{3mm}  receive $\langle \textsc{GetReply} \ d\rangle$

\STATE \hspace{3mm} $DS_c \leftarrow max (DS_c, d.ds)$

\RETURN $d.v$

\STATE \vspace{1mm} \textbf{PUT} (key $k$, value $v$)
\STATE \hspace{3mm} $i = L(k)$ 
\STATE \hspace{3mm} send $\langle \textsc{PutReq} \ k,v,DS_c \rangle$ to server $i$
\STATE \hspace{3mm} receive $\langle \textsc{PutReply} \ tg, ut \rangle$ 
\STATE \hspace{3mm} $DS_c \leftarrow max(DS_c, \langle tg, ut\rangle)$

}
\end{algorithmic}
}

\end{algorithm}

To read the value of an object, the client calls GET method with the desired key to read. The client also specifies the id of the checking group that the server must use. We will see how the server uses this id in Section \ref{sec:serverSide}. We find the preferred server to read the object using the given load balancer service $L$. After finding the preferred server to ask for the key, we send a \textsc{GetReq} request to the server. In addition to the key and the checking group id, we include the client dependency set $DS_c$ in the request message. The server tries to find the most recent version that is consistent by the client's past reads. In the Section \ref{sec:serverSide}, we explain how the server looks for a consistent version based on the $DS_c$.

To write a new value for an object, the client calls PUT method. The server writes the version and records client's $DS_c$ as the dependency of the version. After receiving a response from the server for a GET (or PUT) operation, we update $DS_c$ such that any later version written by the client depends on the version read (or written) by this operation.


\subsection{Server-side}
\label{sec:serverSide}

 In this section, we focus on the server-side of the protocol. We denote the physical clock at server $i$ by $PC_i$. To satisfy $v_1 \deps v_2 \Rightarrow v_1.t > v_2.t$ condition, and assign timestamps close to the physical clocks, \name relies on Hybrid Logical Clocks (HLCs) \cite{hlc}.  $HLC_i$ is the value of HLC at server $i$.  Each server keeps a version vector that has one entry for each tracking group denoted by  $VV_i$. $VV_i[t]$ is the minimum of latest timestamps that server $i$ has received from servers in tracking group $t$. To keep each other updated, servers send heartbeat messages to each other in case of not sending any replicate message for a specific amount of time.
 If there is no key that is hosted by both server $i$ and a server in tracking group $t$, then $VV_i[t] = +\infty$. Each server is a member of one or more dependency checking groups. Servers inside a checking group, periodically share their VVs with each other and compute Stable Version Vector (SVV) as the entry-wise minimum of VVs. 
$SVV^{cg}_i$ is the SVV computed in server $i$ for checking group $cg$.

Algorithm \ref{alg:server1} shows the algorithm for PUT and GET operations at the server-side. 
When a client asks to read an object, the server waits if there exists $\langle t, h\rangle$ in $ds$ such that $VV_i[t] < h$ which means the server is not updated enough, and reading from the current version chain can violate causal consistency. When the server made sure for any  $\langle t, h\rangle$ in $ds$, $VV_i[t] \geq h$, it checks the $SVV^{cg}_i$. If for any  $\langle t, h\rangle$ in $ds$, $SVV^{cg}_i[t] \geq h$, the server returns the most recent version for $k$ such that for any  $\langle t, h\rangle$ in $k.ds$, $SVV^{cg}_i[t] \geq h$. This guarantees that the client never has to wait if it only reads from servers in checking group $cg$. If a client uses different checking groups for different reads, it is possible that the server finds  $\langle t, h\rangle$ in $ds$, such that $SVV^{cg}_i[t] < h$. In this situation, server forgets about $SVV^{cg}_i$, and gives the client the most recent version that has for $k$. Note that this version is guaranteed to be causally consistent with client's previous reads. 

Once server $i$ receives a PUT request, the server updates $HLC_i$ by calling $updateHLC(dt)$ where $dt$ is the highest timestamp in $ds$. Next, the server creates a new version for the key specified by the client.
The server updates $VV_i[T(i)]$ with the new $HLC_i$ value, and sends back its tracking group, $T(i)$, and the assigned timestamp, $d.ds[T(i)]$, to the client in a $\textsc{PutReply}$ message. 

Upon creating a new version for an object in one server, we send the new version to other servers hosting the object via replicate messages. Upon receiving a $\langle \textsc{Replicate} \ k, d\rangle$ message from server $j$, the receiving server $i$ adds the new version to the version chain of the object with key $k$. The server also updates the entry for server $T(j)$ in its version vector (i.e., $VV_i[T(j)]$).  

\begin{algorithm} 
{
\caption{PUT and GET operations at server $i$}
\label{alg:server1}
\begin{algorithmic} [1]
{\footnotesize
\INPUT Tracking grouping function $T$, Data placement function $H$

\STATE \textbf{Upon} receive $\langle \textsc{GetReq} \ k, cg, ds\rangle$

\STATE \hspace{3mm} \textbf{while} there is a member $\langle t, h\rangle$ in $ds$, $h > VV_i[t]$
\STATE \hspace{6mm} wait

\STATE \hspace{3mm} \textbf{if} for all $\langle t, h\rangle$ in $ds$, $h > SVV^{cg}_i[t]$
\STATE \hspace{6mm} \label{line:get:obtain} $d = $ latest version $d$ from version chain of key $k$  \\
\hspace{6mm} s.t.  for any member $\langle t, h\rangle$ in $k.ds$, $h \leq SVV^{cg}_i[t]$
\STATE \hspace{3mm} \textbf{else} 
\STATE \hspace{6mm}  \label{line:get:obtain} $d = $ latest version $d$ from version chain of key $k$  
\STATE \hspace{3mm} send $\langle\textsc{GetReply} \  d\rangle$ to client

\STATE \vspace{1mm} \textbf{Upon}  receive $\langle \textsc{PutReq} \ k, v, ds \rangle$
\STATE \hspace{3mm} \label{line:takeMaxforDt} $dt \leftarrow$ maximum value in $ds$ 
\STATE \hspace{3mm}  $updateHCL(dt)$
\STATE \hspace{3mm}  Create new item $d$
\STATE \hspace{3mm}  $d.v, d.ds \leftarrow v, max(ds, \langle T(i), HLC_i\rangle)$
\STATE \hspace{3mm}  insert $d$ to version chain of $k$
\STATE \hspace{3mm} update $VV_i[T(i)]$ with $HLC_i$
\STATE \hspace{3mm}  send $\langle \textsc{PutReply} \ T(i), d.ds[T(i)]\rangle$ to client
\STATE \hspace{3mm}  \textbf{for} each server $j \neq i$, such that $j \in H(k)$ 
\STATE \hspace{6mm}  \label{line:sendReplicate} send $\langle \textsc{Replicate} \ k, d \rangle$ to server $j$

\STATE \vspace{1mm} \textbf{Upon}  receive $\langle \textsc{Replicate} \ k, d\rangle$ from server $j$
\STATE \hspace{3mm}  insert $d$ to version chain of key $k$
\STATE \hspace{3mm} update $VV_i  [T(j)]$ with $d.ds[T(j)]$

\STATE \vspace{1mm} \textbf{updateHLCforPut ($dt$)}
\STATE \hspace{3mm} $l' \leftarrow HLC_i.l$
\STATE \hspace{3mm} $HLC_i.l \leftarrow max (l', PC_i, dt.l)$
\STATE \hspace{3mm} \textbf{if} $(HLC_i.l = l' = dt.l)$ 
\STATE \hspace{6mm} $HLC_i.c \leftarrow max(HLC_i.c ,dt.c)+1$
\STATE \hspace{3mm} \textbf{else if} $(HLC_i.l = l') \  \ HLC_i.c \leftarrow HLC_i.c + 1$
\STATE \hspace{3mm} \textbf{else if} $(HLC_i.l = l) \  \ HLC_i.c \leftarrow dt.c + 1$
\STATE \hspace{3mm} \textbf{else} $HLC_i.c \leftarrow 0$
}
\end{algorithmic}
}
 
\end{algorithm}
 \vspace*{-3mm}
 
 \subsection{Evaluation}

We have implemented \name using DKVF \cite{dkvf}. You can find our implementation of \name in DKVF repository \cite{dkvfRep}. In this section, we provide the results of $2\times2$ and $4\times1$ groupings for applications $App_1$ and $App_2$ explained in Section \ref{sec:intro}. We run the system explained in Section \ref{sec:intro} consisting of $A_1$, $A_2$, $B_1$, and $B_2$ on different data centers of Amazon AWS \cite{aws}. Note that since we focus on partial replication, there is no assumption about $A_1$ and $B_1$ (respectively $A_2$ and $B_2$) to be collocated.


\textbf{Observations for $App_1$. }
$App_1$ consists of two clients $C1$ and $C2$. $C1$ writes the value $0$ using $A_1$. $C2$ reads $0$ (from $A_2$) and writes $1$ (to $A_2$). Subsequently, $C1$ waits to read $1$ and writes $2$ and so on. The best scenario for this case is when you have only two partitions $A_1$ and $A_2$ in the system. Hence, we normalize the throughput with respect to this. 

\begin{figure}[t]
  \centering
  \subfigure[Chaging the location of $B_1$\label{fig:app1Real}]{\includegraphics[scale=0.4]{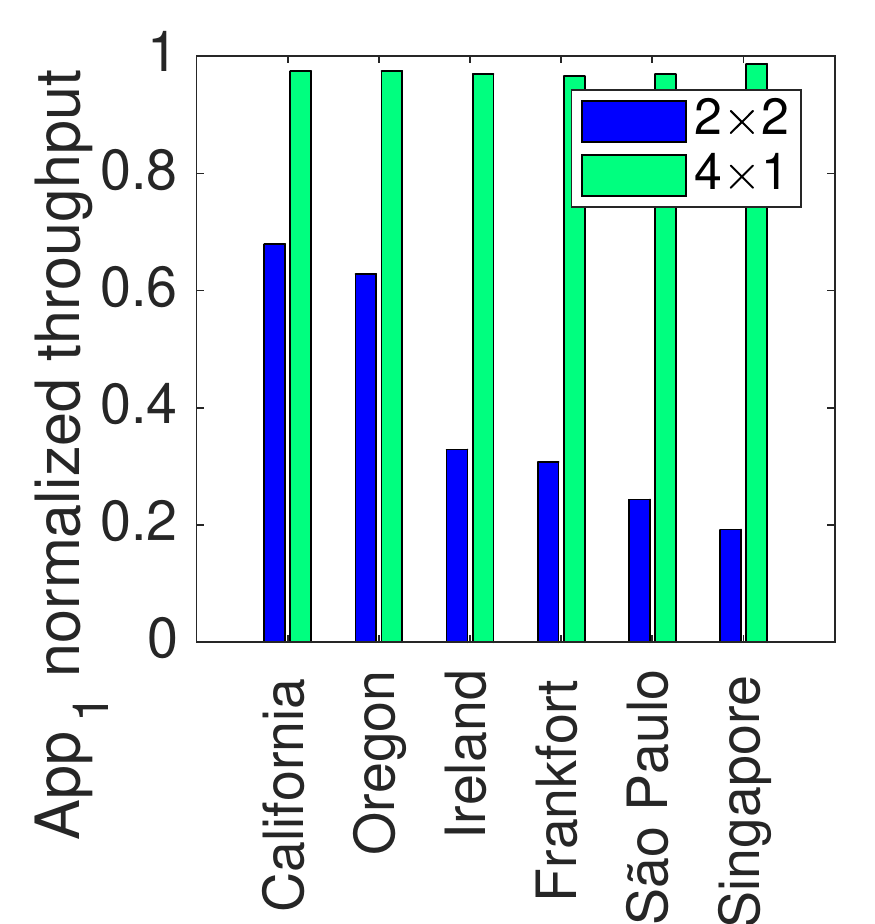}}\quad
  \subfigure[Changing  the delay of $B_1$\label{fig:app1Sim}]{\includegraphics[scale=0.4]{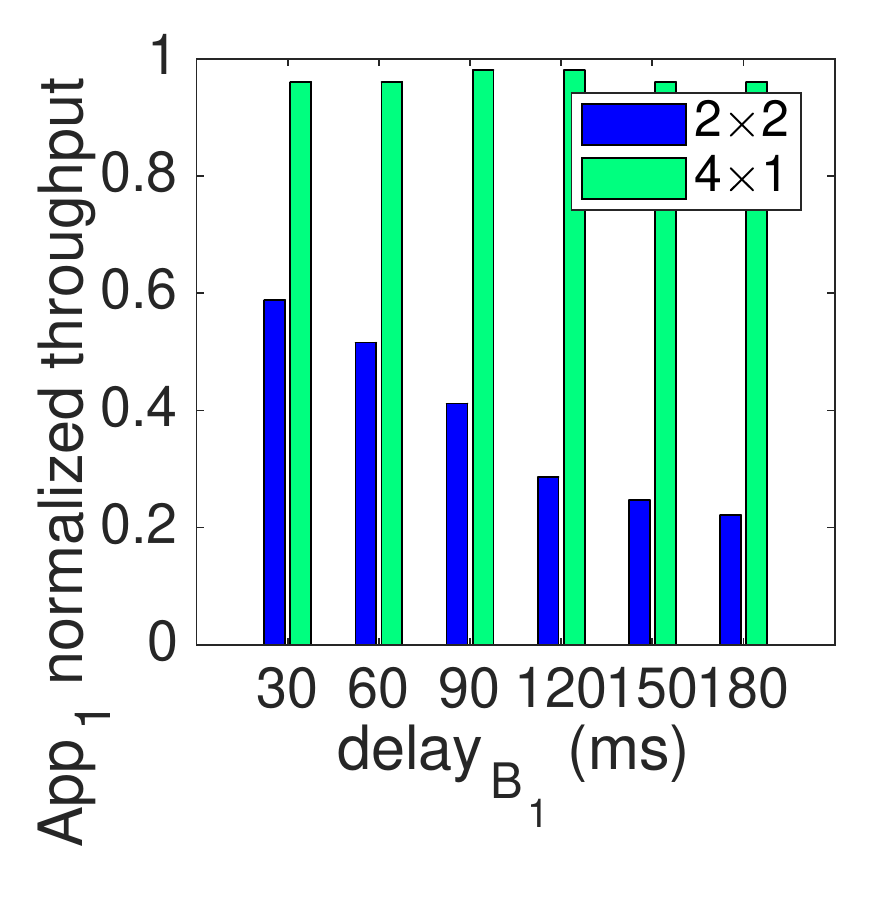}}
   \subfigure[Changing the delay of $B_2$ ]{\includegraphics[scale=0.4]{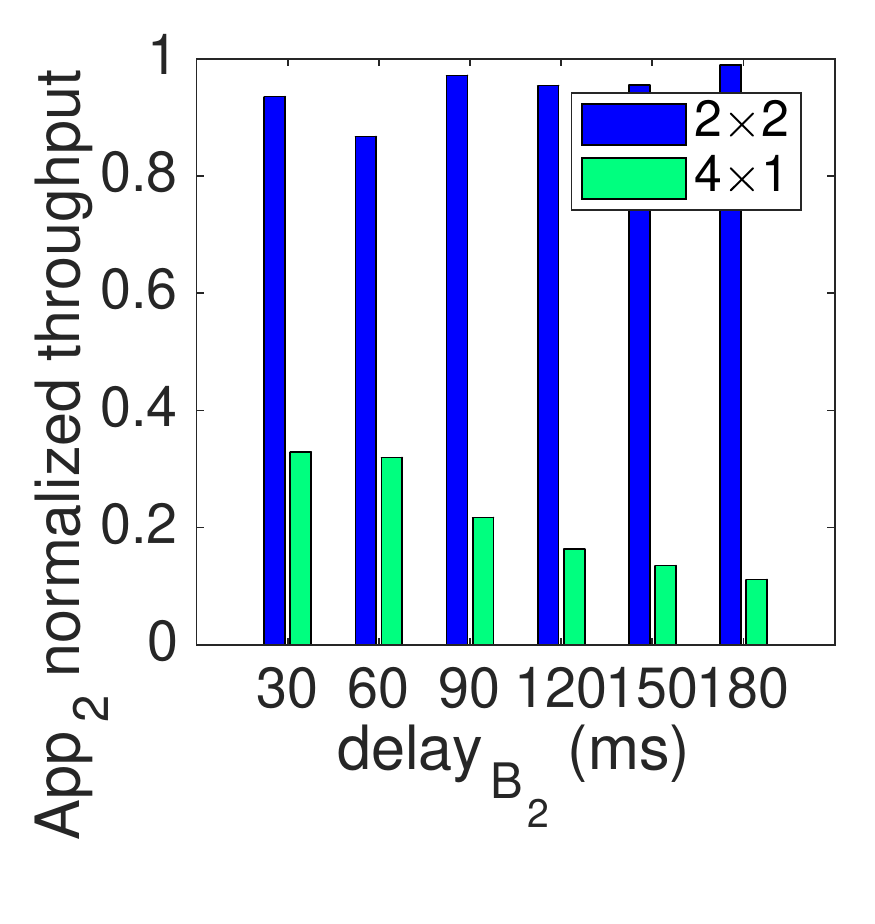}}
 \caption{Normalized throughput of $App_1$ and $App_2$ for different groupings. $App_1$ has higher throughput with $4\times1$, while $App_2$ has higher throughput with $2\times2$.}
 \label{fig:results}
 \vspace*{-5mm}
\end{figure}

The results for $App_1$ are  shown in Figure \ref{fig:results}, where $2\times2$ (respectively $4\times1$) corresponds to the organization in Figure \ref{fig:full} (respectively, Figure \ref{fig:partial}).  In Figure \ref{fig:app1Real}, locations of $A_1$, $A_2$ and $B_1$ are fixed, and we vary the location of $B_1$ from California to Singapore (ordered based on increasing ping time from $A_1$ located in California).
In Figure \ref{fig:app1Sim}, we keep the location of $B_1$ fixed, but artificially add $delay_{B_1}$ to any message sent by $B_1$. As we can see, by viewing the system as Figure \ref{fig:partial}, $App_1$ performance is unaffected whereas viewing the system as Figure \ref{fig:full}, performance drops by more than 50\%.

\vspace*{-0.25mm}
\textbf{Observations for $App_2$. }
In $App_2$, client $C3$ alternates reading from $A_1$ and $B_1$. To provide fresh copies, another client writes the same objects on $A_2$ and $B_2$ respectively. Here, viewing the system as in Figure \ref{fig:partial} drops the performance substantially as the message delay of $B_2$ ($delay_{B_2}$) increases. This is due to blocking the GET operations while waiting for receiving consistent versions. By contrast, by viewing replicas as in Figure \ref{fig:full}, performance remains unaffected. Throughputs are normalized with respect to the case where there is no update.

\section{Discussion}
\label{sec:discussion}


%

In this section, we discuss how our approach differs from existing work on adaptive causal systems and identify future work that we are currently pursuing. 



\textbf{Dynamically adding or removing checking groups:}
Adding checking groups is a straightforward process. Each checking group is associated with a data structure (e.g., SVV in the algorithm provided in Section \ref{sec:protocol}) that the servers need to maintain (in RAM).  Hence, if we want to add a new checking group, the system can run the protocol to initialize these fields and make the new checking group available. 
Removing a checking group is somewhat challenging especially if some client is using it. In this case, we anticipate that the principle-of-locality would be of help. If a client has not utilized a checking group for a while, in most cases,  all the data the client has read has been propagated to all copies in the system. In other words, if a client is using a checking group that has disappeared, we can have the client choose a different checking group. It is unlikely to lead to delays, as all replicas already have the data that the client has read. 
Two practical questions in removing checking groups are (1) the \textit{time} after which we can remove a checking group and (2) how servers can determine that no client has accessed that checking group in that time. 
A more difficult question in this work is \textit{when} to add a new checking group and \textit{how many} checking groups to maintain. Clearly, we cannot create a checking group for each possible client, as it would require exponentially many checking groups. 


\textbf{Utilizing multiple checking groups simultaneously:} \ 
Yet another question is whether clients could have multiple checking groups or whether clients can change their checking group. The former would be desirable when the system does not offer a checking group that the client needs. However, the client could choose two (or few) checking groups whose union is a superset of the checking group requested by the client. In this case, the server providing the data would have to utilize all of these checking groups --on the fly-- to determine which data should be provided to the client.


\textbf{Learning required checking groups automatically:}
In this case, the system will learn from client requests to identify when new checking groups should be added and when existing checking groups should be removed. 
 We expect that dynamically changing the checking groups in this manner would be beneficial due to principle-of-locality, where clients are likely to access data that \textit{similar} to the data they accessed before.  (Recall that we assume that keys are partitioned with semantic knowledge rather than by approaches such as uniform hashing).  
We anticipate that learning techniques such as evolutionary or machine learning techniques would be useful to identify the checking groups that one should maintain.  
 



\textbf{Dynamically changing the tracking groups:} 
Dynamically changing the tracking groups is more challenging but still potentially feasible in some limited circumstances. The reason for this is that while checking groups affect the data maintained by the servers at run-time (in RAM) tracking groups affect storage affected by keys (in long-term permanent storage). In other words, at runtime, we may run into a key that was stored with a different tracking groupings. In this case, it is necessary to convert the data stored with the old tracking grouping into the corresponding data in the new tracking grouping. We expect that principle-of-locality would be of help in this context as well; keys stored long ago are likely to have been updated in all replicas. Conversion of the data stored with keys is protocol specific but still feasible. For example, if we wanted to switch between tracking grouping used by CausalSpartan \cite{causalSpartan} (where a vector DSV is maintained with one entry per replica) to GentleRain \cite{gentleRain} (where only a scalar entry GST is maintained) then we could convert the DSV entry into a GST entry that corresponds to the minimum of the DSV entries. However, the exact approach to do this for different tracking groupings requires semantic knowledge of those tracking groupings. 

\textbf{Comparison with related work:}
Our approach for providing adaptivity in causal consistency is different from other approaches considered in the literature. 
%
Occult \cite{occult} utilizes structural compression to reduce the size of the timestamps. Other approaches include bloom filters \cite{chainReaction}. 
While these features are intended as a configurable parameter, we believe that it is not possible to dynamically change it at run-time while preserving causal consistency (or detection of its violation). 
Furthermore, in all these cases, the reconfiguration provided is client-agnostic; it does not take client requests into consideration. By contrast, our framework provides the ability to allow different clients a view of the system in a manner that improves their performance. Finally, it is possible to take client requests into consideration to identify how adaptivity should be provided.

\section{Conclusion}
\label{sec:con}
%

In this paper, we focused on developing a system that provides causal consistency in an adaptive manner. Specifically, we introduced the notion of tracking and checking groups as a way to generalize existing protocols as well as to develop new adaptive protocols. We provided a framework that, unlike existing causal consistency protocols, can be configured to work with different tracking and checking groupings. This flexibility enables us to trade off between conflicting objectives, and provide different views to different applications so that each application gets the best performance. We argue that the approach and the framework introduced in this paper provide a basis for adaptive causal consistency for replicated data stores.

\newpage
\clearpage
\bibliography{bib}

\begin{thebibliography}{10}

\bibitem{aws}
Amazon aws. \url{https://aws.amazon.com/}.

\bibitem{dkvfRep}
{DKVF}. \url{https://github.com/roohitavaf/DKVF}.

\bibitem{facebook}
Phillipe Ajoux, Nathan Bronson, Sanjeev Kumar, Wyatt Lloyd, and Kaushik
  Veeraraghavan.
\newblock Challenges to adopting stronger consistency at scale.
\newblock In {\em HotOS}, 2015.

\bibitem{chainReaction}
S{\'e}rgio Almeida, Jo{\~a}o Leit{\~a}o, and Lu{\'\i}s Rodrigues.
\newblock Chainreaction: a causal+ consistent datastore based on chain
  replication.
\newblock In {\em Proceedings of the 8th ACM European Conference on Computer
  Systems}, pages 85--98. ACM, 2013.

\bibitem{bost}
Bernadette Charron-Bost.
\newblock Concerning the size of logical clocks in distributed systems.
\newblock {\em Information Processing Letters}, 39(1):11--16, 1991.

\bibitem{okapi}
Diego Didona, Kristina Spirovska, and Willy Zwaenepoel.
\newblock Okapi: Causally consistent geo-replication made faster, cheaper and
  more available.
\newblock {\em arXiv preprint arXiv:1702.04263}, 2017.

\bibitem{orbe}
Jiaqing Du, Sameh Elnikety, Amitabha Roy, and Willy Zwaenepoel.
\newblock Orbe: Scalable causal consistency using dependency matrices and
  physical clocks.
\newblock In {\em Proceedings of the 4th Annual Symposium on Cloud Computing},
  SOCC '13, pages 11:1--11:14, New York, NY, USA, 2013.

\bibitem{gentleRain}
Jiaqing Du, C\u{a}lin Iorgulescu, Amitabha Roy, and Willy Zwaenepoel.
\newblock Gentlerain: Cheap and scalable causal consistency with physical
  clocks.
\newblock In {\em Proceedings of the ACM Symposium on Cloud Computing}, SOCC
  '14, pages 4:1--4:13, New York, NY, USA, 2014.

\bibitem{hlc}
Sandeep~S Kulkarni, Murat Demirbas, Deepak Madappa, Bharadwaj Avva, and Marcelo
  Leone.
\newblock Logical physical clocks.
\newblock In {\em International Conference on Principles of Distributed
  Systems}, pages 17--32. Springer, 2014.

\bibitem{lamport}
Leslie Lamport.
\newblock Time, clocks, and the ordering of events in a distributed system.
\newblock {\em Commun. ACM}, 21(7):558--565, July 1978.

\bibitem{cops}
Wyatt Lloyd, Michael~J. Freedman, Michael Kaminsky, and David~G. Andersen.
\newblock Don't settle for eventual: Scalable causal consistency for wide-area
  storage with cops.
\newblock In {\em Proceedings of the Twenty-Third ACM Symposium on Operating
  Systems Principles}, SOSP '11, pages 401--416, New York, NY, USA, 2011.

\bibitem{eiger}
Wyatt Lloyd, Michael~J Freedman, Michael Kaminsky, and David~G Andersen.
\newblock Stronger semantics for low-latency geo-replicated storage.
\newblock In {\em NSDI}, volume~13, pages 313--328, 2013.

\bibitem{occult}
Syed~Akbar Mehdi, Cody Littley, Natacha Crooks, Lorenzo Alvisi, Nathan Bronson,
  and Wyatt Lloyd.
\newblock I can't believe it's not causal! scalable causal consistency with no
  slowdown cascades.
\newblock In {\em NSDI}, pages 453--468, 2017.

\bibitem{causalSpartan}
Mohammad Roohitavaf, Murat Demirbas, and Sandeep Kulkarni.
\newblock Causalspartan: Causal consistency for distributed data stores using
  hybrid logical clocks.
\newblock In {\em Reliable Distributed Systems (SRDS), 2017 IEEE 36th Symposium
  on}, pages 184--193. IEEE, 2017.

\bibitem{gentleRain+}
Mohammad Roohitavaf and Sandeep Kulkarni.
\newblock Gentlerain+: Making gentlerain robust on clock anomalies.
\newblock {\em arXiv preprint arXiv:1612.05205}, 2016.

\bibitem{dkvf}
Mohammad Roohitavaf and Sandeep Kulkarni.
\newblock Dkvf: A framework for rapid prototyping and evaluating distributed
  key-value stores.
\newblock {\em arXiv preprint arXiv:1801.05064}, 2018.

\end{thebibliography}

\end{document}